\documentclass{mpe_report}

\usepackage{psfig,graphicx,epsfig}
\usepackage{color}
\usepackage{amsmath,amssymb,epic,eepic,array}

\begin{document}

\pagenumbering{arabic}
\setcounter{page}{24}

\renewcommand{\FirstPageOfPaper }{ 24}\renewcommand{\LastPageOfPaper }{ 27}

\title{Recent Observations of $EGRET$ Pulsar Wind Nebulae}
\author{Mallory S.E. Roberts\inst{1} \and E. V. Gotthelf\inst{2} \and Jules P. Halpern\inst{2} \and
Crystal L. Brogan\inst{3} \and Scott M. Ransom\inst{3}}  
\institute{Eureka Scientific, Inc. 2452 Delmer St. Suite 100, Oakland, CA 94602-3017, USA \and
Columbia University, Pupin Hall, 550 West 120th St, New York, New York 10027 USA \and
NRAO, 520 Edgemont Rd.  Charlottesville, VA  22901 USA}
\maketitle

\begin{abstract}
We present recent X-ray and radio observations of pulsar wind nebulae discovered
in $EGRET$ error boxes. Two XMM-Newton observations show the X-ray
extent of the rapidly moving PWN associated with the variable $\gamma$-ray source 3EG J1809$-$2328, 
and a trail coming from the new millisecond pulsar PSR J1614$-$2230
at high Galactic z. We also briefly discuss three PWN that are $HESS$ TeV 
sources including a new $HESS$ source
we argue is associated with the Eel nebula in 3EG J1826$-$1302.
\end{abstract}

\section{Introduction}

$EGRET$, the high-energy ($E > 100$~MeV) instrument on board the Compton Gamma-Ray Observatory, 
localized  $\sim 100$ $\gamma$-ray sources to within $\sim 1.5^{\circ}$ which appear to be  associated with our Galaxy
(\cite{hbb+99}, but see also Grenier et al. (2006), these proceedings). 
Around 20 pulsar wind nebulae (PWN) are known to be coincident with 
these $EGRET$ error boxes (\cite{rbg+05}).
Some are around the known $\gamma$-ray pulsars, while others are around likely
$\gamma$-ray pulsars that have been discovered since 1999 when the $EGRET$ mission ended (\cite{rob05}).
Since both PWN and $\gamma-$ray pulsations are signs of efficient particle acceleration in the 
magnetosphere, it could be expected that a bright $\gamma$-ray pulsar would also have a bright nebula
(for a recent review of pulsars and pulsar wind nebulae, see \cite{krh06}).
However, there are several apparently 
variable $EGRET$ sources whose error boxes contain bright pulsar wind nebulae
as well (\cite{rrk01,ntgm03}). This suggests that sometimes 
$\gamma$-rays come from the nebula, not the pulsar magnetosphere (\cite{rgr02}). 

In section 2. we present recent multi-wavelength observations of nebulae associated 
with three variable $EGRET$ sources. In section 3 we report apparently 
extended  X-ray emission from a millisecond pulsar in a mid-latitude $EGRET$ errror box.

\section{High-Energy emission from RPWN}

Several Galactic $EGRET$ error boxes whose error boxes contain PWN have 
$> 100$~MeV $\gamma$-ray emission which shows evidence of variability at 
the $> 90\%$ confidence level (according to the $V_{12}$ statistic
of Nolan et al. 2003). All of these sources have PWN whose X-ray and 
radio shapes appear to be determined by rapid (i.e. supersonic) 
motion (RPWN, \cite{rbg+05}). 
The location and shape of the shock in the forward direction of these RPWN 
are most likely determined by a balance between  
ram pressure and the pulsar's wind pressure. Hence 
the shock acceleration site
is highly dependent on the density and structure of the pulsar's varying environment. 
Numerical modeling of these systems suggests a narrow trailing wind termination 
shock within a broader, bow-shaped shock
region (\cite{bad05,rah06}). However, these simulations have only just begun 
to model the expected synchrotron emission
and need to take into account
doppler boosting and spectral evolution of the flows as well as the density 
and magnetic field strength (\cite{dvab06}). 
Numerical investigations of the effects of clumpiness or a density gradient 
in the surrounding medium still need to be done. 

Three of these RPWN possibly associated with variable $EGRET$ sources 
were first discovered by X-ray imaging of the $EGRET$ error box with
$ASCA$ (\cite{rrk01}). These images showed extended 
areas of hard X-ray emission. Short Chandra images and follow-on
radio imaging revealed their RPWN nature. Here we discuss recent X-ray and 
TeV images of these sources and their relation to the radio images. 

\subsection{Taz: The Canonical GeV Emitting RPWN?}

3EG J1809$-$2328 has the highest $V_{12}=3.92$ (99.988\% probability of a 
non-constant flux) of any 
non-AGN source. It also has one of the 
best constrained error boxes of any of the unidentified $EGRET$ sources. 
Oka et al. (1999)\nocite{okn+99} noted the ASCA source seemed morphologically related to 
the Lynds 227 dark cloud. From their CO maps they estimated a kinematic
distance of 1.7 kpc to the cloud. They suggested the $\gamma$-rays were a result
of relativistic bremsstrahlung from high-energy electrons generated by the
pulsar interacting with matter in the cloud. However, they assumed
the entire cloud as the target, and it is unclear how that might lead to
emission variable on a time scale of months.
VLA and Chandra imaging of the ASCA source confirm the PWN nature of the 
northern half of the ASCA source while the southern part was resolved into a 
stellar cluster associated with the nearby S32 HII region (\cite{rrk+01,brrk02}). 
An X-ray point source with a short ($\sim 15^{\prime \prime}$) 
trail was revealed in the Chandra ACIS image at the tip of a funnel shaped
radio nebula. We refer to this as the ``Taz" PWN since it looks like the 
pulsar is generating a whirlwind as it travels. Although the X-ray and radio 
spectra and morphologies confirm the PWN nature of Taz, very deep pulse searches
with the Greenbank Radio Telescope have been unsuccessful at discovering a 
radio pulsar.

A 7200~s R band image of the field made at the MDM observatory with the 
1.3~m McGraw-Hill telescope shows the dark cloud (Fig1a). 
The edge of the cloud is clearly delimited by $8.3\mu$ mid-infrared 
emission observed by the MSX mission (Fig. 1, contours). Taz was imaged
with XMM-Newton on 2-3 Oct 2004 for 69~ks with the EPIC MOS instruments
(the PN was in small window mode in order in order to perform a search for 
pulsations which was unsuccessful). The image shows a larger, broader 
X-ray emission region that is similar to what is seen in the Mouse 
(\cite{gvc+04})
and the nebula in the thermal composite SNR IC443 (\cite{gcs+06}). 

\begin{figure}
\centerline{\psfig{file=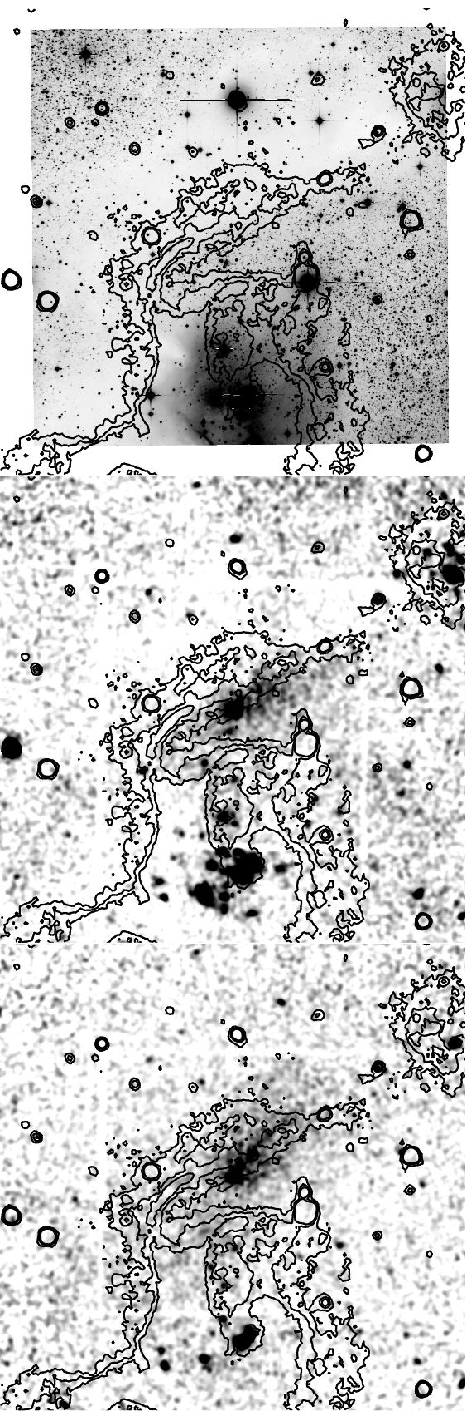,width=6.0cm,clip=} }
\caption{a. (top) R band image of Taz region, with MSX $8.3\mu$ mid-IR contours
showing relationship of mid-IR to the Lynds 227 dark cloud. b. (middle) 
XMM-Newton EPIC MOS1+MOS2 soft (0.5-2.5~keV) X-ray image with MSX contours.
c. (bottom) XMM-Newton hard (2.5-8~keV) X-ray image. Note how the mid-IR 
emission outlining Lynds 227 correlates with the soft X-rays but not the hard. 
\label{fig1}}
\end{figure}
                 
In Fig. 1b and Fig. 1c we show the background subtracted, exposure 
corrected soft (0.5-2.5 keV) and hard (2.5-8 keV) images from the combined 
MOS1 and MOS2 data with the MSX contours. The soft X-ray nebula is clearly
being absorbed by Lynds 227, but there is little evidence that the 
morphology of the hard nebula is affected by the cloud. On the side of the
nebula which is not absorbed by the cloud, the soft X-ray emission appears
a bit broader than the hard emission, suggesting synchrotron burn-off 
perpendicular to the apparent direction of motion. In Fig. 2 we 
overlay the 20~cm radio contours on the hard image. There is a ridge of
emission along the symmetry axis of the radio emission. Overall, Taz appears
to have the small ``tongue" (seen in the Chandra image but largely 
confused with the point source in the EPIC image)  plus broader ``tail" 
morphology which seems to be fairly standard for the X-ray emission of RPWN.
  
\begin{figure}
\centerline{\psfig{file=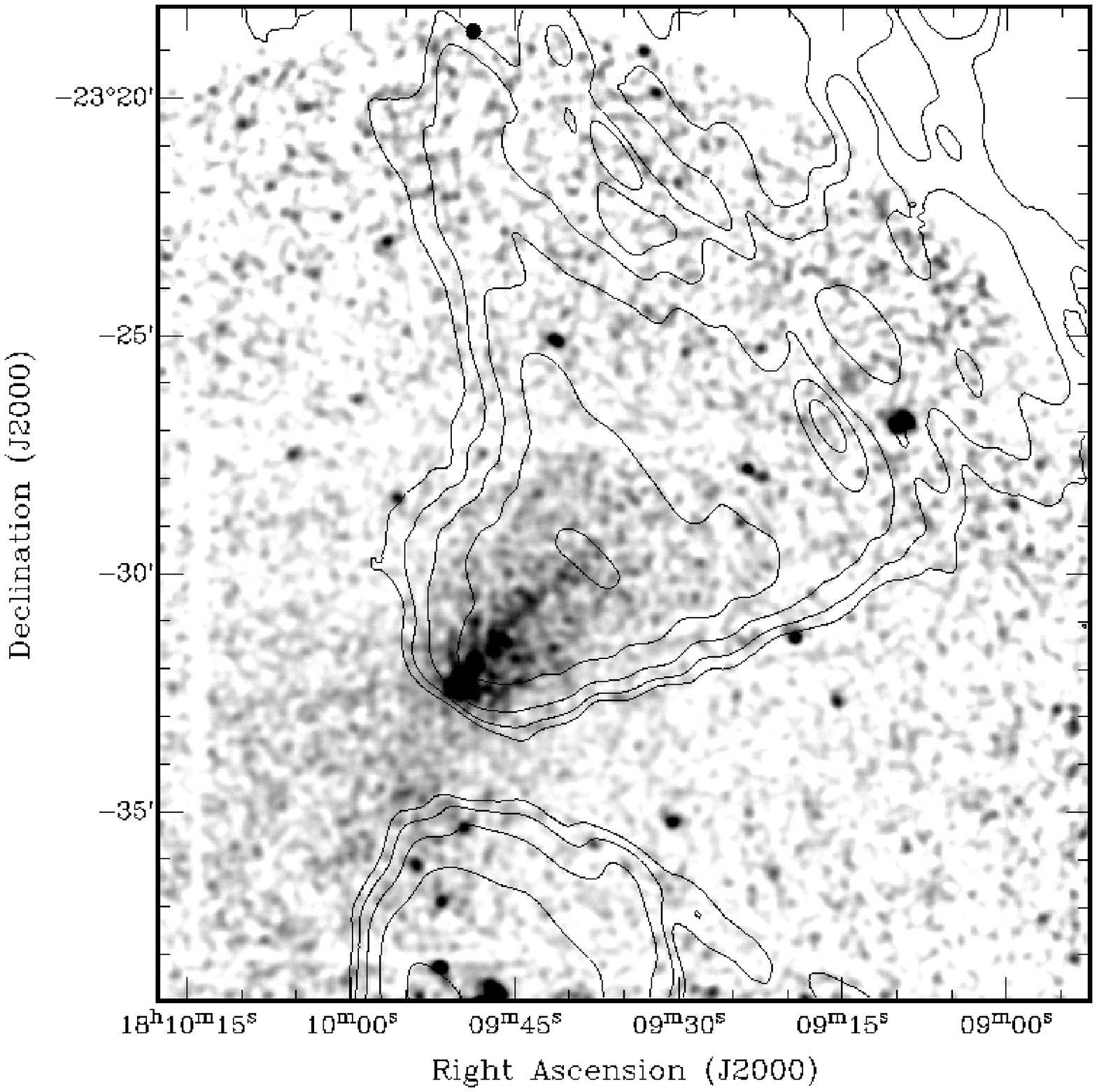,height=6.5cm}}
\caption{XMM-Newton hard (2.5-8~keV) X-ray image of the Taz RPWN in 
3EG J1809$-$2328 with VLA 20~cm radio contours overlain. 
\label{fig2}}
\end{figure}

\subsection{The Eel and Rabbit: GeV - TeV emitting RPWN?}

GeV J1825$-$1310/3EG J1826$-$1302 has a $V_{12}=3.22$, the
second highest of unidentified low-latitude sources. It is
in the region where there is a bright, variable, unidentified Comptel source.
VLA 90~cm and 20~cm images of the region around 3EG J1826$-$1302 
reveal at least four SNR, as well as many molecular clouds and other structures. 
There is also the energetic young pulsar PSR B1823$-$13 just outside
the 95\% error contour (as determined by Roberts, Romani \& Kawai 2001)
of the GeV source. Within the error box, ASCA discovered a $> 15^{\prime}$
nebula surrounded by fainter, diffuse emission. 
A 15~ks exposure with the Chandra ACIS-I array on 17-18 Feb 2003 
resolved part of the nebula into a stellar cluster, but also 
discovered a point-like source with a faint, remarkably long 
($\sim 4^{\prime}$) trail  of hard ($> 2$~keV)
X-ray emission (Fig. 3). There is also excess radio
emission in the 90~cm image whose morphology 
suggests it may be related and does not appear to have associated mid-IR 
emission suggesting it is non-thermal. Because of the long, thin nature 
of the nebula and the difficulty in distinguishing it from its surroundings,
we refer to this as the Eel nebula. 

The HESS collaboration recently reported on a spectral gradiant 
in the extended TeV source that is coincident with PSR B1823$-$13
(\cite{aha06a}). This strongly suggests the pulsar is moving in a northerly 
direction, leaving behind a trailing high-energy nebula. However,
the deep TeV image clearly shows significant excess emission 
to the north of the pulsar, inconsistent with this picture. 
The excess is roughly triangular in shape and we indicate its 
approximate position in Fig. 3. We suggest that this emission is
a separate source which we refer to as HESS J1826$-$131 and is
associated with the Eel nebula, not PSR B1823$-$13. 

\begin{figure}
\centerline{\psfig{file=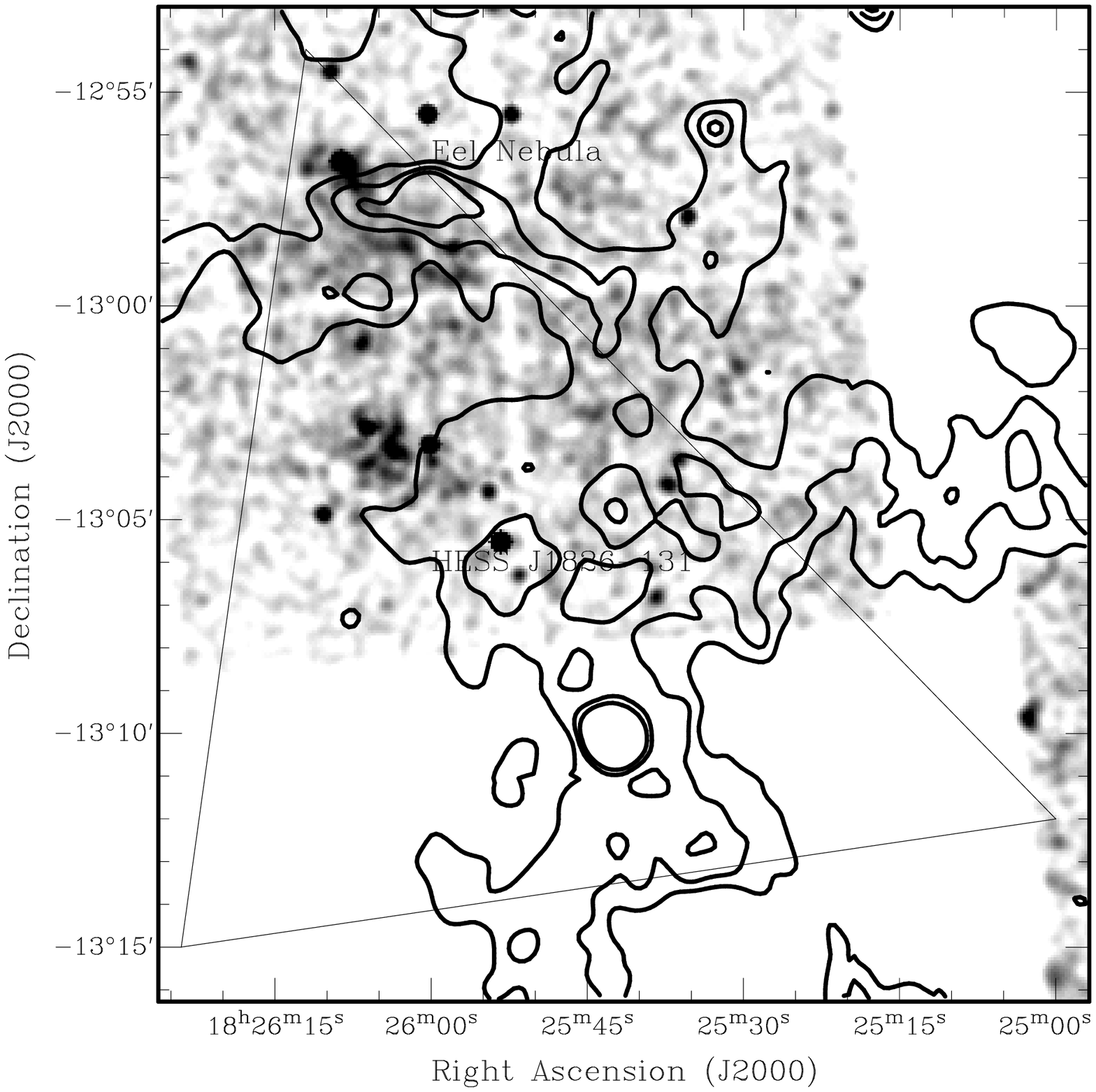,height=6.5cm }}
\caption{Chandra ACIS-I 0.3-8 keV image of the Eel nebula with
90~cm VLA contours. The triangle indicates the rough position
of the new TeV source, HESS J1826$-$131.
\label{fig3}}
\end{figure}

HESS has also discovered two sources in the wings of the Kookaburra radio
complex (\cite{aha06b}) which is in the error box of the
apparently variable ($V_{12}=1.59$) source GeV J1417$-$6100/3EG J1420$-$6038.  
In the Northern wing, the HESS source is between the energetic
young pulsar PSR J1420$-$6048 in the southeast corner of the 
radio wing and a region of polarized radio emission in the
northern half of the wing, apparently confirming the PWN 
nature of the wing. However, the morphology of the X-ray nebula
surrounding the pulsar as revealed by Chandra (\cite{nrr05}) 
is consistent with a slowly moving pulsar that may have previously been 
south of its current position. This suggests the radio nebula may have
been offset by the supernova remnant reverse shock.

The TeV source in the southern wing is between the Rabbit radio
nebula (\cite{rrjg99}) and another region of polarized radio emission
in the wing, suggesting a physical connection between the two (Fig. 4).  
Although the X-ray morphology around the apparent pulsar in the Rabbit is a bit 
uncertain, it is suggestive of an RPWN
with the apparent direction of motion coming from the 
wing.  If the Rabbit and the wing are one nebula,
the overall radio morphology is 
reminiscent of the Mouse which has a bright radio head and a 
fainter radio body (\cite{gvc+04}). 

\begin{figure}[h!]
\centerline{\psfig{file=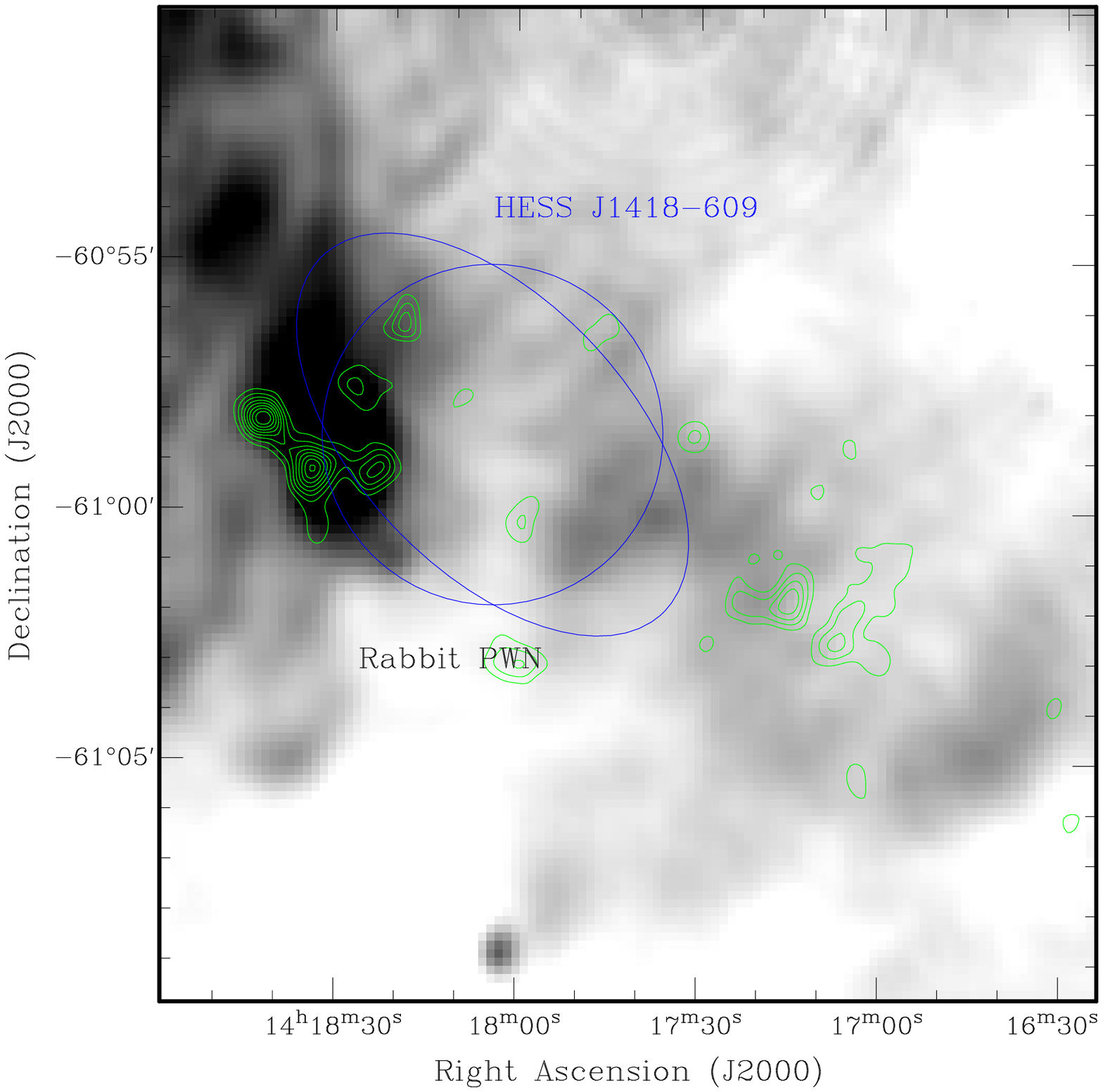,height=6.8cm} }
\caption{20~cm radio image of Rabbit and lower wing of the Kookaburra. 
The contours show regions of linearly polarized radio emission. The circular
and elliptical fits to the extended HESS TeV source are also shown.
\label{image}}
\end{figure}

\section{PSR J1614$-$2230: A New RPWN Around a Millisecond Pulsar}

PSR J1614$-$2230 is an energetic 3.15~ms pulsar recently discovered in
a survey of mid-latitude $EGRET$ sources with the Parkes
radio telescope (\cite{hrr+05}).
It is unique among fast millisecond pulsars in the Galactic field in
having a companion whose mass is $> 0.4 M_{\odot}$. It is in 
the error box of the apparently non-variable source 3EG J1616$-$2221
which is one of the few mid-latitude sources that has survived the
reanalysis of the $EGRET$ data by Grenier et al. (these proceedings). 
If we assume the pulsar is the $\gamma$-ray source then the 
the inferred $\gamma$-ray efficiency is $\sim 8\%$ 
assuming the dispersion measure distance of 1.3~kpc and 1 steradian
beaming. This, as well as the $EGRET$ spectrum $\Gamma = 2.42\pm 0.24$, 
is similar to that of the millisecond pulsar
PSR J0218+4232 which has a claimed, albeit marginal, pulse detection
in the $EGRET$ data (\cite{khv+00}). If PSR J1614$-$2230 is a $\gamma$-ray 
source, we might expect X-ray emission from the pulsar, either from
heated polar caps or magnetospheric pulsations.

XMM-Newton observed PSR J1614$-$2230 for $\sim 5$~ks on 17 Aug 2005. 
The combined PN plus MOS image (Fig. 5) reveals a thermal ($kT \sim 0.2$~keV)
point source coincident with the pulsar. However, there is also hard X-ray 
emission to the southwest of the pulsar which does not appear 
to be point-like. If this is a trailing PWN, the implied
direction of motion is consistent with a preliminary proper motion 
measurement obtained from $\sim 3.5$ years of timing data with Parkes
and the Greenbank Telescope. 
Given the nominal distance, the height above the Galactic plane
is $d_z \sim 450$~pc or about three times the scale height of the
HI disk. This pulsar therefore may provide a unique probe of the ISM at 
high Galactic z.

\begin{figure}
\centerline{\psfig{file=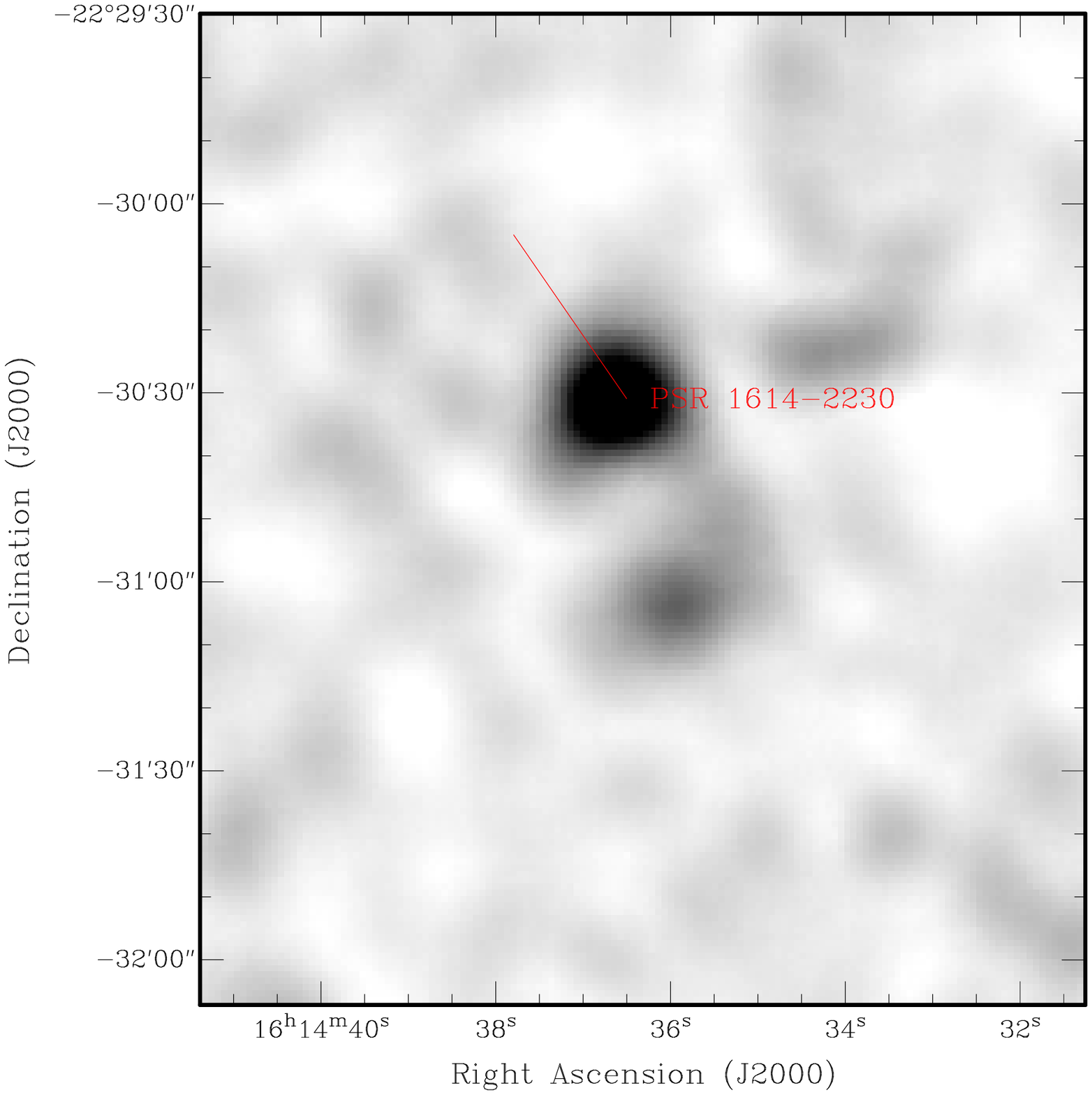,height=7.0cm} }
\caption{XMM-Newton 0.3-5 keV X-ray image of PSR J1614$-$2230 
from the combined EPIC PN, MOS1 and MOS2 data. The preliminary 
estimate of the direction of proper motion from timing data (which still has
significant uncertainty) is indicated.
\label{fig5}}
\end{figure}

\vskip 0.4cm

\begin{acknowledgements}
This work made use of the Australia Telescope which is funded by the 
Commonwealth of Australia for operation as a National Facility managed by
CSIRO; the National Radio Astronomy Observatory which is a facility of the 
National Science Foundation operated under cooperative agreement by 
Associated Universities, Inc.; XMM-Newton, an ESA science mission with 
instruments and contributions directly funded by ESA Member States and NASA;
Chandra X-ray Observatory which is operated by the 
Smithsonian Astrophysical Observatory for and on behalf of NASA.
\end{acknowledgements}
   


          \clearpage

\end{document}